\documentclass[12pt]{elsarticle}

\journal{Ultramicroscopy}

\biboptions{authoryear}

\usepackage{amssymb,amsmath}
\usepackage{graphicx}
\renewcommand{\vec}[1]{\boldsymbol{#1}} 
 
\begin{document}

\begin{frontmatter}

\title{A new solution to the curved Ewald sphere problem for 3D image reconstruction in electron microscopy}

\author{J. P. J. Chen}
\author{K. E. Schmidt}
\author{J. C. H. Spence}
\author{R. A. Kirian}
\address{
Department of Physics\\
Arizona State University\\
Tempe, AZ 85287 USA}

\begin{abstract}
We develop an algorithm capable of imaging a three-dimensional object given a collection of two-dimensional images of that object that are significantly influenced by the curvature of the Ewald sphere. These two-dimensional images cannot be approximated as projections of the object. Such an algorithm is useful in cryo-electron microscopy where larger samples, higher resolution, or lower energy electron beams are desired, all of which contribute to the significance of Ewald curvature.
\end{abstract}


\end{frontmatter}

\section{Introduction}
In this paper we propose an iterative projection algorithm to reconstruct a 3-dimensional (3D) object from 2-dimensional (2D) real-space images in the case where the depth-of-field is less than the thickness of the sample, so that these 2D images are not projections of the sample scattering density.  This is also referred to as the ``Ewald sphere curvature problem" in the context of cryo-electron microscopy (cryo-EM) \cite{Wolf_2006,Leong_2010,Russo_2018,Zhu_2018} and it becomes important for resolutions and incident wavelengths where the thickness of the sample exceeds the depth-of-focus of the lens system \cite{Spence_2013}.  
In that case, bright-field transmission electron microscope (TEM) images no longer represent valid 2D projections of the object, so that conventional cryo-EM software (based, for example, on the method of filtered back-projection) cannot be used to merge the 2D images into a 3D reconstruction without losses in resolution. Unlike some of the previous approaches discussed below, our method does not rely on application of a corrected transfer function, but builds in the relevant scattering theory from the outset, and inverts this from all the data using the technique of iterative projection algorithms (IPAs).  A solution to the Ewald curvature problem would allow 3D reconstructions from lower beam energies, higher resolutions, and with thicker samples, all factors of which contribute to the severity of the Ewald curvature.

For a scattering vector $\vec{q}=\vec{k}-\vec{k}_0$, the Ewald sphere is defined as the spherical surface traced out by $\vec{k}$ under elastic scattering conditions, so that negligible energy is transferred to the sample. Here $\vec{k}_0$ is the incoming wavevector of the electron beam, while $\vec{k}$ is the scattered wavevector.  The Ewald curvature problem arises when images are formed from a range of $\vec q$-vectors over which the Ewald sphere cannot be well-approximated as a plane. In that case, the Fourier-projection theorem, relating planes in reciprocal space to projections of the object in real space, as assumed in many current reconstruction and data merging algorithms, cannot be used.  For a crystal (where $\vec{q}$ extends to a reciprocal lattice point, defining a Bragg condition), a curved sphere means that beams scattered at equal angles to the direct beam on opposite sides of the diffraction pattern are no longer Friedel pairs represented by conjugate complex structure factors. The two different complex structure factors (four real numbers) are combined by the lens to form one set of interference fringes (defined by just two quantities) contributing to the image.  In previous work by \cite{Wolf_2006,Russo_2018}, a correction algorithm was described that relies on the fact that (for a crystal), Bragg beams (whose width is limited to that of a small particle) separate laterally as they progress downstream. This is seen in far-out-of-focus shadow images of small crystals, which separate into one shadow image for each Bragg beam. At resolution $d$, the separation due to defocus, $D_f$, alone is  $2 D_f \lambda/d$ (other aberrations may also contribute) \cite{Zuo_Spence_2017}.  This allows the interferences contributed by one side of the diffraction pattern to be separated from those due to the other. \cite{Russo_2018} then generalize their method to non-periodic samples with continuous scattering, and apply their corrections using an elegant segmented transfer function divided into wedges. 
By comparison, our method does not require the use of images recorded at large defocus or additional images which may introduce radiation damage. However our method proposed here is much more computationally intensive, and may be well suited to images formed with the use of a Zernike phase plate.

The conventional formulation of bright-field, weak-phase object, high-resolution transmission electron microscopy imaging assumes either a sample transmission function for a phase object (based on an eikonal approximation) using a projected potential \cite{Spence_2013}, or uses the first Born approximation for electron scattering.  The phase shift introduced by the sample is assumed to be less than ninety degrees. 
 However, a new formulation is needed for the case of a curved sphere, since the image is then formed from  scattering on the Ewald sphere rather than onto a plane in reciprocal space, and a projected potential cannot be used. 
As detailed in Section~\ref{sec_Model}, we use the standard Born approximation for the wave function, but we do not make a far-field approximation. Therefore our method gives the correct relationship between the phase of the direct and scattered radiation \cite{Lentzen_2014}.   
As for the case of optical microscopy where the Ewald sphere curvature causes the depth-of-field to be smaller than the sample thickness (as used for the optical sectioning technique), it is then necessary to compute wavefields on successive planes normal to the beam within the sample, and to propagate these onward to an area detector, or to the ``exit face" plane across the downstream face of the sample. Focus adjustment then makes it possible to ``look inside" a semi-transparent object albeit with limited depth-of-focus and artifacts caused by the other planes of the sample. For modern aberration-corrected transmission electron microscopes, this depth-of-field can be as small as a few nanometers. The validity domain of the approximations used in this paper therefore limit our method to weakly scattering objects for which multiple scattering is negligible but Ewald sphere curvature is significant. As discussed elsewhere in the context of the inversion of multiple scattering \cite{Donatelli_Spence_2020}, multiple scattering may also occur with a flat Ewald sphere for samples consisting of heavy elements.

Recently, \cite{Gureyev_etal_2019a,Gureyev_etal_2019b} gave a comprehensive treatment of the Ewald curvature problem and proposed a solution method they call ``pattern matching tomography." In a separate work, \cite{Ren_etal_2020} proposed an iterative strategy based on finding a 3D potential that best fits the measured intensities with regularization, for inorganic non-periodic samples. They also incorporate the multislice algorithm, allowing their technique to account for multiple scattering. 
The algorithm we propose here is similar in nature to the method developed by \cite{Ren_etal_2020}, with the difference being that we use ``projection operators" (operations that make the minimum change to an input) to seek out a 3D potential that fits the measured intensities, as opposed to cost function minimization by calculating gradients.
Iterative procedures similar in spirit of our work have been proposed previously for example by \cite{Allen_etal_2004} where they reconstruct 2D objects (exit wave functions) from a through-focus series of 2D images.

This paper is structured as follows: Section~\ref{sec_Model} introduces the mathematical model that we will use to describe the imaging process and our criterion for the Ewald curvature to become significant. Section~\ref{sec_Algorithm} outlines our proposed algorithm.
Section~\ref{sec_Simulation} shows the results of simulations with implementation details, before drawing conclusions in Section~\ref{sec_Conclusion}.

\section{The model}
\label{sec_Model}
The objective of our work is to reconstruct a 3D image of an object given a collection of 2D images formed with the object in different orientations and located at different distances from the focal plane. Assuming no multiple scattering (the first-order Born approximation), ideal plane-wave illumination, and an ideal imaging system, the measurement from the experiment can be calculated as follows. 

Starting with the Schr\"odinger equation as detailed in \ref{appendix_LS}, it can be shown that the scattered wavefront $\psi(\vec{r})$ from an incoming plane wave incident upon an object described by a potential energy distribution $V(\vec{r})$ under the Born approximation is given by 
\begin{align}
\psi(x,y,z) 
= e^{ik_0 z} \left [ 1
-i \frac{\pi}{E\lambda} \int d^3r'
\int \frac{dk_x}{2\pi} \int \frac{dk_y}{2\pi}
e^{ik_x (x-x')}
e^{ik_y (y-y')}
e^{i\left (\sqrt{k_0^2-k_x^2-k_y^2}-k_0\right)(z-z')}
V(\vec{r}') \right ] \,,
\end{align}
where $\vec{r} = (x,y,z)$, $z>z'$, $k_x$ and $k_y$ are the spatial frequency components along the transverse directions $x$ and $y$, $k_0 = 2\pi / \lambda$, $E$ and $\lambda$ are the energy and wavelength of the incident plane wave, respectively.
For $z<z'$ a virtual image is produced.
Introducing the scaled object potential
\begin{equation}
f(\vec{r}) = \frac{\pi}{E\lambda} V(\vec{r})
\end{equation}
and the wavevector transfer 
\begin{align}
 \vec{q}
&= (q_x, \; q_y, \; q_z)
\nonumber
\\
&= \vec{k} - \vec{k}_0 
\nonumber
\\
&= (k_x, \; k_y, \; k_z - k_0) \;,
\end{align}
where $\vec{k} = (k_x, \; k_y, \; k_z)$ is the outgoing wavevector, $\vec{k}_0 = (0, \; 0, \; k_0)$ is the incoming wavevector, $k_z = \sqrt{k_0^2 - k_x^2 - k_y^2}$, and $q_z = k_z - k_0$.
The scattering intensity, $| \psi(x,y,z) |^2$, can be put into the form  
\begin{align}
\big | \psi(x,y,z) \big|^2
&= \left | 
1 
- i \int dz'
\int \frac{dq_x}{2\pi} \int \frac{dq_y}{2\pi}
e^{i(q_x x + q_y y)}
\;
e^{i q_z(z-z')}
\int {dx'} \int {dy'}
e^{-i(q_x x'+q_y y')}
f(\vec{r}') \right|^2 \;.
\label{eq_scat_int}
\end{align}
Equation~(\ref{eq_scat_int}) can be written as 
\begin{align}
\big| \psi(x,y,z) \big|^2
&= \left | 
1 
- i \int dz'
\mathcal{F}^{-1}_{\perp}
\;
e^{i q_z(z-z')}
\mathcal{F}_{\perp}
f(\vec{r}') \right|^2 \;,
\label{eq_psi_sq}
\end{align}
where $\mathcal{F}_{\perp}$ and $\mathcal{F}_{\perp}^{-1}$ are the 2D Fourier transform and the inverse 2D Fourier transform, respectively, in the transverse directions $x$ and $y$ to the propagation axis.
Expanding equation~(\ref{eq_psi_sq}) gives
\begin{align}
\big| \psi(x,y,z) \big|^2
&= 
1 
+
2\; \mathrm{Im} \left( \int dz'
\mathcal{F}^{-1}_{\perp}\;
e^{i q_z(z-z')}
\mathcal{F}_{\perp}
f(\vec{r}')
 \right)
+ 
\left | \int dz'
\mathcal{F}^{-1}_{\perp}\;
e^{i q_z(z-z')}
\mathcal{F}_{\perp}
f(\vec{r}') \right|^2
 \;.
\label{eq_psi_sq_expanded}
\end{align}
Since the Born approximation only keeps terms linear in the potential,
the quadratic term in the potential in equation~(\ref{eq_psi_sq_expanded}) will be correct only at points where the cross term between the direct and
scattered waves is negligible. Because we are in the bright-field
regime, we must either drop the second order term or include the second
Born approximation in this cross term.
In typical bright-field electron microscopy, the second order term in $f(\vec{r})$ is much smaller than unity and thus we choose to ignore it here. We also drop the constant bright-field term in equation~(\ref{eq_psi_sq_expanded}) under the assumption that it can be removed at an early stage in the data processing. Thus the 2D images obtained in a bright-field electron microscope in our model are proportional to
\begin{align}
\mathbb{I}_z(x,y)
&= 
\mathrm{Im} \left( \int dz'
\mathcal{F}^{-1}_{\perp}\;
e^{i q_z(z-z')}
\mathcal{F}_{\perp}
f(\vec{r}')
 \right) \; ,
\label{eq_I_continuous}
\end{align}
where $\mathbb{I}_z(x,y)$ is understood as the continuous 3D scattering intensity evaluated at a specific value of $z$. That value of $z$ is referred to here as the ``focal distance'' that defines the focal plane of the image.

Equation (\ref{eq_I_continuous}) is the governing equation that we use in this paper to describe the electron microscopy imaging process. 
We do not incorporate the effect of the lens transfer function in this current work, however we present a derivation of it now for context and completeness. Define the scattering angle $\theta$ as the angle between $\vec{k}$ and $\vec{k}_0$ (see Figure~\ref{fig_EwaldCurvature}). For $\theta$ small, $q_z$ can be approximated as
\begin{align}
q_z &= k_0(\cos \theta - 1) 
\approx - \frac{2 \pi}{\lambda} \frac{\theta^2}{2}
\\
&= k_z - k_0
\approx - \frac{\lambda}{4 \pi} \left(q_x^2 +q_y^2\right) \;.
\end{align} 
With the inclusion of spherical aberration $C_s$, the scattering intensity becomes
\begin{align}
\mathbb{I}_z(x,y)
&= 
\mathrm{Im} \left( \int dz'
\mathcal{F}^{-1}_{\perp}\;
\exp\Big({-i \chi(q_x,q_y,\Delta f)}\Big)
\mathcal{F}_{\perp}
f(\vec{r}')
 \right) \; ,
\label{eq_I_continuous_with_CTF}
\end{align}
where $\Delta f = z - z'$ is the deviation from Gaussian focus and
$\chi(q_x,q_y,\Delta f)$ is the familiar lens transfer function
\begin{align}
\chi(q_x,q_y,\Delta f) 
&= \frac{\lambda}{4 \pi} \left(q_x^2 +q_y^2\right) \Delta f  
- 2{C_s} \left(\frac{\lambda}{4\pi}\right)^3 \left(q_x^2 +q_y^2\right)^2
\nonumber
\\
&= \frac{2\pi}{\lambda} 
\left(\frac{1}{2} \theta^2 \Delta f 
- \frac{1}{4}C_s \theta^4 \right) \;.
\label{eq_chi_CTF}
\end{align}
As mentioned above, our proof-of-principle work presented here in this paper do not consider the lens transfer function and we work with equation~(\ref{eq_I_continuous}) instead of (\ref{eq_I_continuous_with_CTF}).

\begin{figure}[!ht]
\centering
\includegraphics[scale=0.4]{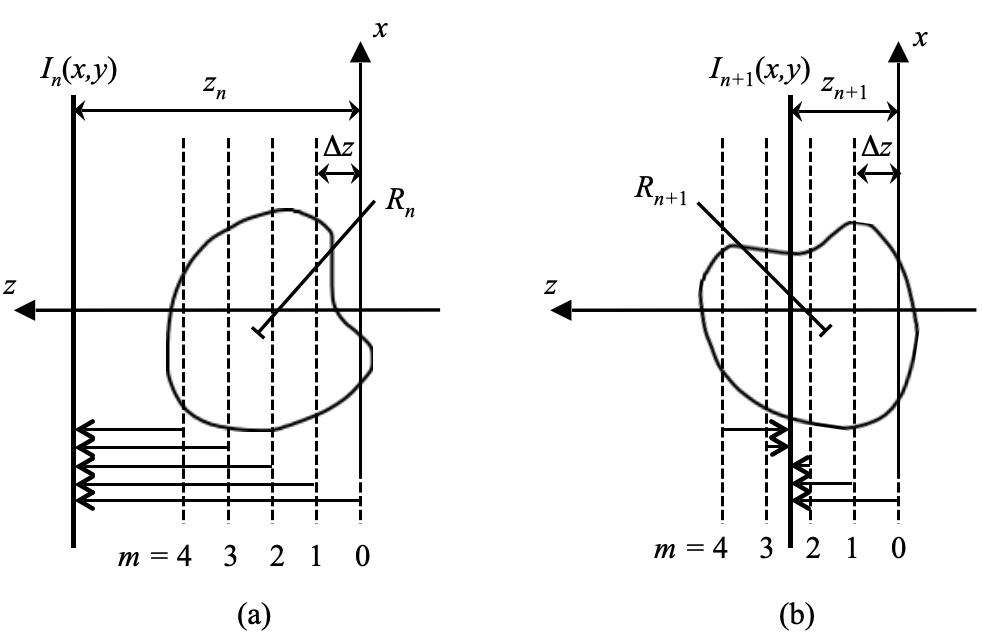}
\caption{Two schematic examples of how views are calculated. Wavefronts emanating from sections of the object (dashed line) separated by $\Delta z$ and indexed by $m$ are propagated to a common plane (thick solid line) a distance $z_n$ along the $z$ axis from a chosen origin and coherently summed. The imaginary part of the resultant 2D image, $I_n(x,y)$, is what is referred to as a ``view." Note the two examples have the object in different orientations ($R_n$ and $R_{n+1}$) and different focal distances ($z_n$ and $z_{n+1}$).}
\label{fig_exampleViews}
\end{figure}

In order to simulate equation~(\ref{eq_I_continuous}) on a computer, we discretize $f(\vec{r})$ into a total of $M$ 2D sections along the direction of propagation, each separated by a distance $\Delta z$. Then we may write
\begin{align}
\mathbb{I}_z(x,y)
\simeq
\mathrm{Im} \left( 
\sum_{m=0}^{M-1} 
\mathcal{F}_{\perp}^{-1} e^{i q_z (z - m \Delta z)} \mathcal{F}_{\perp} 
f(x,y,m \Delta z)
\right) 
\label{eq_I_discrete}
\end{align}
up to a proportionality constant $\Delta z$.
Since in an electron microscopy experiment, many images with the sample in different orientations and different distances from the focal plane are typically measured, 
we write the $n$th measured data image, $I_n(x,y)$, referred to here as a ``view,'' as 
\begin{align}
I_n(x,y)
= 
\mathrm{Im} \left( 
\sum_{m=0}^{M-1}
\mathcal{F}_{\perp}^{-1} e^{i q_z (z_n - m \Delta z)} \mathcal{F}_{\perp} 
f_n(x,y,m \Delta z)
\right) \;,
\label{eq_I_data}
\end{align}
where $z_n$ is the focal distance for the $n$th view and
\begin{equation}
f_n(\vec{r}) = \mathcal{R}_n f(\vec{r}) 
\label{eq_f2f_n_rotation}
\end{equation}
is the scaled object potential rotated by a 3D rotation operator $\mathcal{R}_n$ that rotates a 3D function to the $n$th orientation. 
If we define the free-space propagator as 
\begin{align}
 \mathcal{P}(\zeta) \equiv \mathcal{F}_{\perp}^{-1} e^{i q_z \zeta} \mathcal{F}_{\perp} \; ,
\end{align}
where $\zeta$ is the propagation distance, we may simplify equation~(\ref{eq_I_data}) to
\begin{align}
I_n(x,y)
= 
\mathrm{Im} \left( 
\sum_{m=0}^{M-1}
\mathcal{P}(z_n - m \Delta z)  
f_n(x,y,m \Delta z)
\right) \;.
\label{eq_I_data_L}
\end{align}
The above expressions show that each view can be understood as the imaginary part of a coherent superposition of the scattering amplitudes from a series of thin 2D sections that are all propagated to a common plane. 
This process is depicted schematically in Figure~\ref{fig_exampleViews} for two example views. 
The objective of our work can then be stated as follows: reconstruct the 3D scaled object potential $f(\vec{r})$ given a set of $N$ 2D views, $\{I_n(x,y)\}$. 

\subsection{Criterion for significant Ewald curvature}
The effect of Ewald curvature is contained in the term $q_z m \Delta z$ in the exponential of equation~(\ref{eq_I_data}). 
In terms of more useful quantities, this effect of the Ewald curvature depends on three factors: (1) the wavelength of the incoming plane wave, $\lambda$, (2) the desired resolution, $d$, and (3) the size of the sample itself, $L$. 
For single-particle cryo-EM we assume a roughly spherical particle, where $L$ becomes the propagation distance along the beam path. For a slab-shaped sample, $L$ is the particle thickness.
The relationship between the three factors, $\lambda$, $d$ and $L$ is derived in this section and has been shown previously by \cite{DeRosier_2000, Spence_2013, Downing_Glaeser_2018}.
\begin{figure}[!ht]
\centering
\includegraphics[scale=0.4]{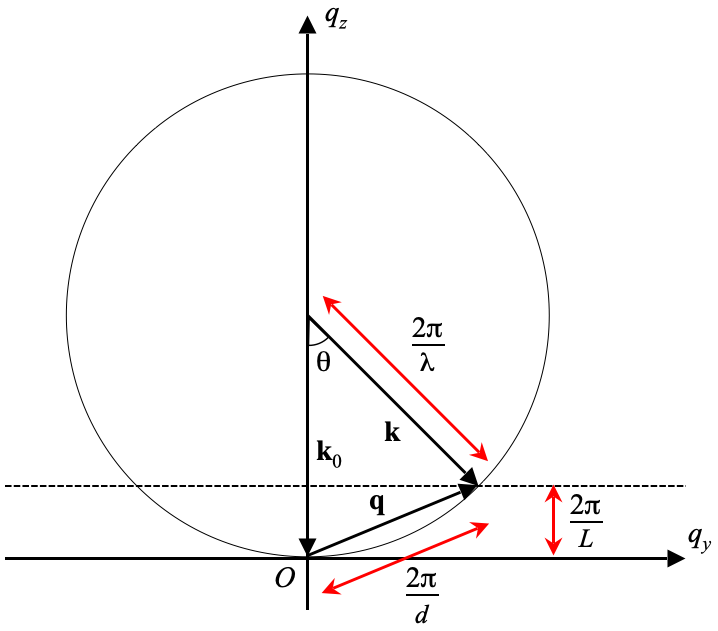}
\caption{Reciprocal space construction for determining the condition for Ewald curvature to become significant. The circle shows the Ewald sphere. The dashed line indicates the position of the first zero of the sinc function described in the main text. $O$ is the origin in Fourier space.}
\label{fig_EwaldCurvature}
\end{figure}

Consider an object of uniform density with thickness $L$ in the $z$ direction. Such an object has the Fourier transform in the $q_z$ direction proportional to 
\begin{align}
F(q_z) &= \int_{-L/2}^{L/2} dz \; e^{i q_z z}
\nonumber
\\
&= L \; \frac{\sin \left(q_z{L}/{2} \right)}{q_z{L}/{2}}.
\end{align}
A microscope lens collects elastic scattering amplitudes that are proportional to this Fourier transform.  For a given view, the amplitudes are constrained to lie on the Ewald sphere, and are limited by the maximum scattering angle that the lens can collect.
The values sampled by the Ewald sphere are indistinguishable from the values sampled on a planar slice through the origin in Fourier space within regions where $F(q_z)$ is relatively constant.
The first zero of $F(q_z)$, in this case a sinc function, occurs at ${q_z L}/{2} = \pi$, so the effect of Ewald curvature is definitely significant when 
\begin{equation}
q_z \geq \frac{2 \pi}{L}.
\label{eq_significant_ewald_curvature}
\end{equation}
Writing $\vec{q}$ in terms of spherical coordinates gives, 
\begin{equation}
\begin{split}
\vec{q} &= k_0 \left (
\sin\theta \cos\phi \; \hat{x}
+ \sin\theta \sin\phi \; \hat{y} 
+ \cos\theta \hat{z} 
-\hat{z} \right)
\\
&= 2k_0\sin(\theta/2)
\Big [
\cos(\theta/2)\cos\phi\hat x
+\cos(\theta/2)\sin\phi\hat y
-\sin(\theta/2)\hat z 
\Big ]
\end{split}
\end{equation}
so that equation~(\ref{eq_significant_ewald_curvature})
becomes
\begin{equation}
L \geq \frac{\lambda}{2\sin^2(\theta/2)}.
\label{eq_L_inequality1}
\end{equation}
With
\begin{align}
q = |\vec{q}| = 2k_0 \sin(\theta/2) = \frac{4\pi \sin(\theta/2) }{\lambda}
\,,
\end{align}
the resolution $d$ defined so that $qd = 2\pi$ is
\begin{align}
d = \frac{\lambda}{2\sin(\theta/2)} \;.
\label{eq_res}
\end{align}
Relating the resolution from equation~(\ref{eq_res}) to (\ref{eq_L_inequality1}) then yields
\footnote{The phase factor in the Fresnel propagator gives $L = {d^2}/({4\pi \lambda})$ which is equivalent to requiring $q_z L \geq 1/4$ instead of $q_z L \geq 2\pi$, as is used in the derivation here for Ewald curvature to become significant.}
\begin{align}
L \geq \frac{2d^2}{\lambda}
\label{eq_DOF}
\end{align}
as the condition for the effect of the Ewald sphere to be significant.
Equation~(\ref{eq_DOF}) is equivalent to the requirement that the Fresnel number $F = d^2/(L \lambda)$ exceeds 0.5 for the effect of the Ewald sphere to be ignored, in agreement with \cite{Downing_Glaeser_2018}.
If $L$ is interpreted as a propagation distance along the beam direction, and $d$ the resolution limit imposed by a lens, then ${2d^2}/{\lambda}$ is the depth-of-field, i.e., the range of planes considered to be in focus at resolution $d$, referred to the object space.
Equation~(\ref{eq_DOF}) therefore expresses the condition that the depth-of-field is less than the sample thickness, so that ``optical sectioning" (with resolution ${2d^2}/{\lambda}$ reckoned along $z$) is possible for 3D imaging by recording images for many values of $z$ in equation (\ref{eq_I_continuous}). For a defocus value of ${2d^2}/{\lambda}$, $d$ also gives the width of the first Fresnel edge fringe, or the width of zones in a Fresnel zone plate with focal length ${2d^2}/{\lambda}$.

\begin{figure}[!ht]
\centering
\includegraphics[scale=0.45]{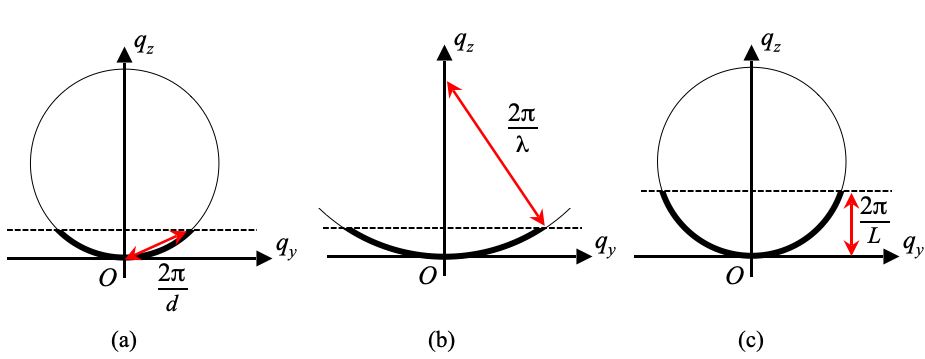}
\caption{The effect of the curvature of the Ewald sphere can be ignored by choosing (a) low enough resolution $d$, (b) small enough wavelength $\lambda$, or (c) small enough object size $L$. The thick solid curve shows the region in Fourier space where the Ewald curvature is considered to be not significant. The dashed line indicates the position of the first zero of the sinc function described in the main text. $O$ is the origin in Fourier space.}
\label{fig_EwaldCurvatureThreeCases}
\end{figure}

Equation~(\ref{eq_DOF}) written in another way becomes
\begin{align}
\varepsilon = \frac{2d^2}{\lambda L} \leq 1,
\label{eq_Ewald_inequality}
\end{align}
where we have defined the dimensionless number $\varepsilon$, which is the ratio of the depth-of-field to the object thickness, and is equal to twice the Fresnel number.
Equation~(\ref{eq_Ewald_inequality}) says: if $\varepsilon$ is less than or equal to one, the effect of Ewald curvature is significant.
Figure~\ref{fig_EwaldCurvature} shows how the parameters $\lambda$, $d$ and $L$ can be related in Fourier space at the condition where the effect of the Ewald sphere becomes significant under our criterion.
Figure~\ref{fig_EwaldCurvatureThreeCases} shows geometrically the effect on the region where the Ewald sphere is considered ``flat" when one of those three parameters is altered while the other two remain fixed.

\begin{figure}[!ht]
\centering
\includegraphics[scale=0.4]{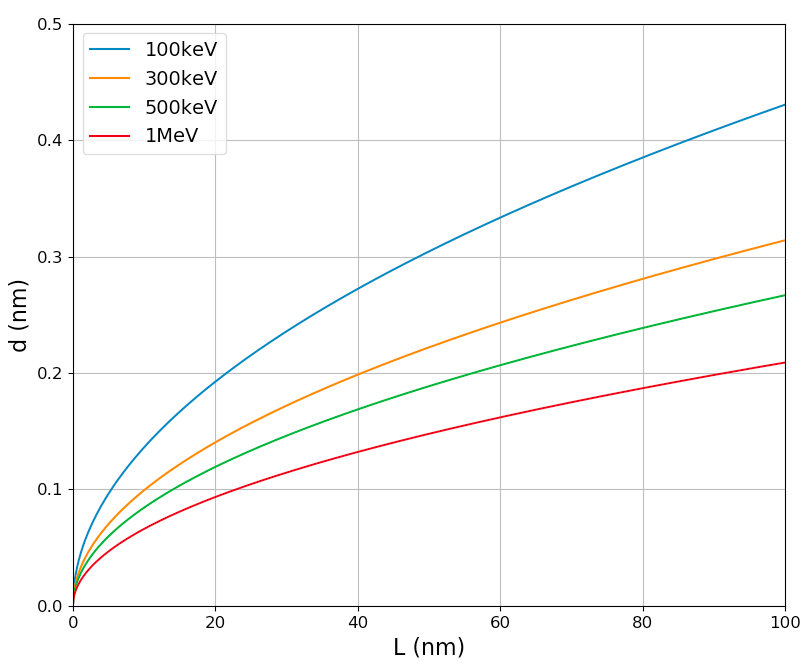}
\caption{Object size $L$ versus resolution $d$ at which Ewald curvature is considered to be significant according to equation~(\ref{eq_Ewald_inequality}) for electrons at four different kinetic energies.}
\label{fig_Lvsres}
\end{figure}

\begin{figure}[!ht]
\centering
\includegraphics[scale=0.5]{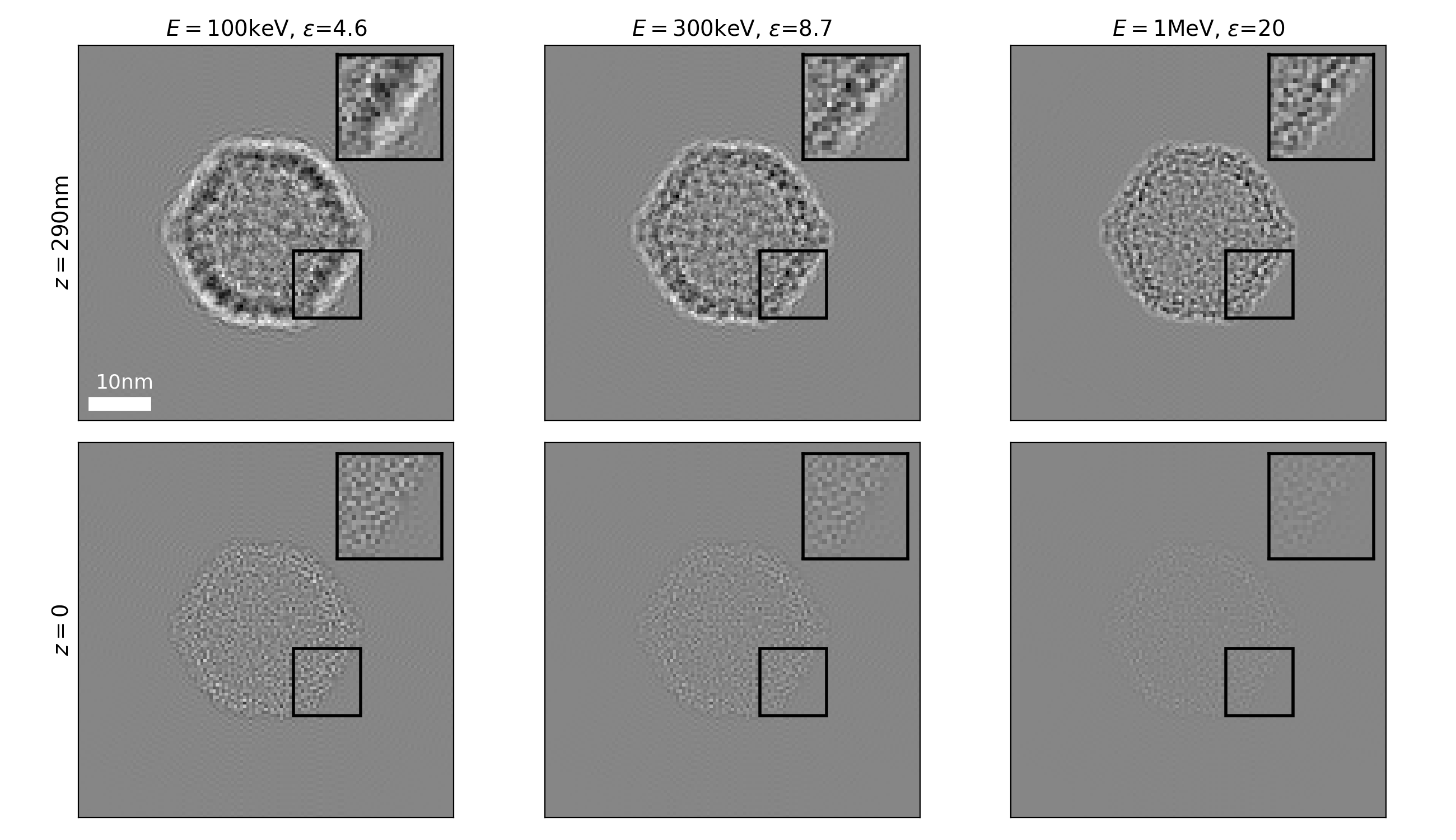}
\caption{Example views of a virus particle at three different electron kinetic energies and two different focal distances. The virus in this simulation is 29nm across. The greyscales are set to the same maximum and minimum values in all subfigures.}
\label{fig_views}
\end{figure}

As an example, for $100$keV electrons ($\lambda \approx 0.037$~\AA{}) and an object of thickness $L=60$nm, the resolution at which Ewald curvature becomes relevant is $d = 3.33$~\AA{}.
Figure~\ref{fig_Lvsres} shows the resolution at which the effect of Ewald curvature becomes significant for different object thicknesses at four different beam energies.
Figure~\ref{fig_views} shows example views of a virus particle calculated at three different incident beam energies and two different focal distances.

\clearpage
\section{Algorithm}
\label{sec_Algorithm}
Many problems can be posed in terms of the satisfaction of multiple constraints. The solution to the original problem requires that all of the constraints be satisfied. The geometric interpretation of this is that the solution is located at the intersection of all surfaces defined by the constraints in a high dimensional space. 
One way to arrive at the intersection, and hence solve the original problem, is via algorithms called iterative projection algorithms (IPAs). 
Operations to ``project" onto individual constraint surfaces, referred to as ``projection operators", are constructed where they make the minimum possible change to an iterate, denoted here by $\mathbf{f}$, such that a specific constraint is satisfied. The boldface indicates that $\mathbf{f}$ is the vectorized representation of a discrete function, but can also be the vectorized representation of a collection of functions in general, as is the case for our proposed algorithm.
The projection operators can be combined to form deterministic rules that update $\mathbf{f}$ in such a way as to progressively satisfy all constraints. 
See \cite{Gerchberg_Saxton_1972,Fienup_1982,Bauschke_etal_2002,Elser_2003,Luke_2005,Elser_etal_2007,Marchesini_2007,Millane_Lo_2013} and references therein for more in-depth discussions on IPAs and constraint-satisfaction problems.

Typically the update rule for the IPA is formulated such that the next iterate is generated from a combination of two projection operators, denoted here by $P_S$ and $P_M$, acting on the current iterate at the $j$th iteration, $\mathbf{f}^{(j)}$. 
The simplest IPA is the Error Reduction (ER) algorithm \cite{Gerchberg_Saxton_1972,Fienup_1982} where the $j$th iterate is updated according to the rule:
\begin{equation}
\mathbf{f}^{(j+1)} = P_S P_M \mathbf{f}^{(j)} \;.
\label{eq:ER}
\end{equation}
The ER algorithm moves the iterate steadily towards a fixed-point but is unable to escape and explore other regions of the multi-dimensional space should that fixed-point turn out not to be a solution. Such outcomes are common when the constraints involved are non-convex in the defined space.
A more effective update rule is the Relaxed-Averaged-Alternating Reflections (RAAR) algorithm \cite{Luke_2005} in which the $j$th iterate is evolved according to the rule:
\begin{equation}
\mathbf{f}^{(j+1)} = 
\beta \mathbf{f}^{(j)}
 - \beta P_S  \mathbf{f}^{(j)} 
+(1 - 2 \beta) P_M \mathbf{f}^{(j)}
+ 2\beta P_S P_M \mathbf{f}^{(j)} 
\; ,
\label{eq:RAAR}
\end{equation}
where $\beta$ is a real-valued parameter of the algorithm. 

The above formulation deals naturally with two constraints. When there are more than two constraints, an effective strategy was proposed by \cite{Gravel_Elser_2008} in an approach they have coined ``divide-and-concur." In this approach the iterate $\mathbf{f}$ contains as many copies of the discrete function $f$ as there are constraints: this is the so-called ``divide" step of the algorithm. In one of the projection operators each copy of $f$ is made to satisfy a single individual constraint (hence the reason for having one $f$ for each constraint), the other projection operator replaces the set of all $f$ with the average $f$, calculated by $\frac{1}{N} \sum_{n=1}^N f_n$ for $N$ equally weighted constraints: this is the so-called ``concurrence" step, which enforces the requirement that all copies of $f$ must describe the same discrete function. 
We shall identify the iterate in the case of $N$ constraints with curly brackets $\mathbf{f} = \{ f_n \}$ where this notation is understood to mean the  iterate $\mathbf{f}$ contains the set of all $N$ $f_n$, and more precisely, $\mathbf{f}$ is the concatenated vectorization of all $N$ discrete functions $f_n$.

\subsection{Our problem}
\begin{figure}[!ht]
\centering
\includegraphics[scale=0.35]{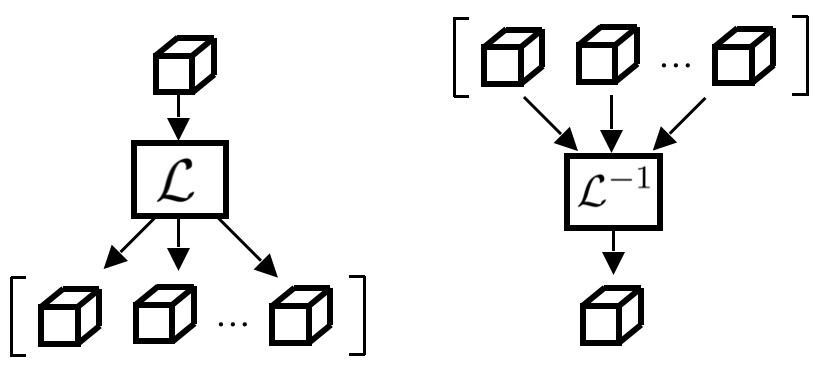}
\caption{The action of $\mathcal{L}$ and its inverse. The 3D arrays each represent an individual $f(x,y,z)$. The square brackets denote the set $\{ f_n(x,y,z) \}$, which is the iterate $\mathbf{f}$ in our iterative projection algorithm.}
\label{fig_algo_L}
\end{figure}

In our problem we have $N$ constraints where each of the constraints corresponds to a single view, $I_n(x,y)$, recorded by the microscope. Therefore, following the divide-and-concur recipe, the iterate $\mathbf{f}$ consists of a set of $N$ volumes $\{ f_n(x,y,z) \}$.
The projections $P_S$ and $P_M$ for our problem then require operators that map a single volume to a set of $N$ volumes, and the corresponding inverse operators that take the set of $N$ volumes and map them back to a single common volume. These operators are defined as follows. 
The forward and inverse propagation operators, as introduced in Section~\ref{sec_Model}, are
\begin{align}
 \mathcal{P}(z) 
 &= 
\mathcal{F}_{\perp}^{-1} e^{i q_z z} \mathcal{F}_{\perp}
\label{eq_propagtor_for}
\\
\mathcal{P}^{-1}(z) 
&= 
\mathcal{F}_{\perp}^{-1} e^{-i q_z z} \mathcal{F}_{\perp}
=
\mathcal{P}(-z) \;.
\label{eq_propagtor_inv}
\end{align}
Define
\begin{align}
\mathcal{L}_{mn} &= \mathcal{P}(z_n - m \Delta z)
\label{eq_op_for}
\\
\mathcal{L}_{mn}^{-1} &= \mathcal{P}^{-1}(z_n - m \Delta z),
\label{eq_op_inv}
\end{align}
so that we may write
\begin{align}
\rho_{mn}(x,y) &= \mathcal{L}_{mn} \; f_n(x,y, m\Delta z)
\\
f_n(x,y, m\Delta z) &=  \mathcal{L}_{mn}^{-1} \; \rho_{mn}(x,y) \;,
\end{align}
where $\rho_{mn}(x,y)$ is the $m$th 2D section from $f(x,y,z)$ rotated by $\mathcal{R}_n$ and propagated to focal distance $z_n$. 
We further define the operators $\mathcal{L}$ and $\mathcal{L}^{-1}$ such that
\begin{align}
\mathcal{L} f(x,y,z) 
&=  \Big\{ \mathcal{R}_n f(x,y,z) \Big\} 
= \Big\{ f_n(x,y,z) \Big\} 
= \mathbf{f}
\\
\mathcal{L}^{-1} \mathbf{f} &= 
\mathcal{L}^{-1} \Big\{ f_n(x,y,z) \Big\} = 
\frac{1}{N} \sum_{n=1}^N \mathcal{R}_n^{-1} f_n(x,y,z) = f(x,y,z) \;.
\end{align}
The operator $\mathcal{L}$ takes $f(x,y,z)$ and makes $N$ copies of it with each copy rotated by $\mathcal{R}_n$ to form the set $\{ f_n(x,y,z) \}$. The operator $\mathcal{L}^{-1}$ takes the set $\{f_n(x,y,z)\}$, inverse rotates each $f_n(x,y,z)$ in that set and outputs the average of those $N$ inverse rotated volumes. The action of $\mathcal{L}$ and $\mathcal{L}^{-1}$ is summarized graphically in Figure~\ref{fig_algo_L}.
Using the defined operators, the $n$th view from equation~(\ref{eq_I_data_L}) can be written as
\begin{align}
I_n(x,y)
= 
\mathrm{Im} \left( 
\sum_{m=0}^{M-1}
\rho_{mn}(x,y)
\right) 
= 
\mathrm{Im} \left( 
\sum_{m=0}^{M-1}
\mathcal{L}_{mn}f_n(x,y,m \Delta z)
\right) 
\;.
\label{eq_I_data_linop}
\end{align}
Note that the information contained in the set of all $MN$ $\rho_{mn}(x,y)$, i.e., $\{ \rho_{mn}(x,y) \}$, is in principle exactly the same as that contained in the set of all $N$ $f_n(x,y,z)$, i.e., $\{ f_n(x,y,z) \}$, due to the unitarity of the forward and inverse propagation operators $\mathcal{P}$ and the rotation operator $\mathcal{R}$. In practice however there will be some loss of information going forward and coming back due to the interpolations required for rotations onto a Cartesian computational grid.

\subsection{Projection Operators}
\label{sec_projectionOps}
Here we define the two projection operators, $P_M$ and $P_S$, for our problem as follows. 
First, denote $\rho_{mn}(x,y)$ at the $j$th iteration of the algorithm by $\rho_{mn}^{(j)}(x,y)$. The iterate at the $j$th iteration can be written similarly as $\mathbf{f}^{(j)}$ and the views given by that iterate as $I_n^{(j)}(x,y)$. The views $I_n^{(j)}(x,y)$ will in general not be equal to the measured data views $I_{n}^{\;\mathrm{data}}(x,y)$ and the projection operator $P_M$ seeks to make them equal with the minimum amount of change to $\rho_{mn}^{(j)}(x,y)$. The appropriate operation can be shown to be (see \ref{appendix_Im_proj} for more details)
\begin{align}
P_{Mn} \; \rho_{mn}^{(j)}(x,y)
=
\rho_{mn}^{(j)}(x,y)
+
i
\frac{1}{M}
\Big( I_{n}^{\;\mathrm{data}}(x,y) - I_n^{(j)}(x,y) \Big) \; ,
\end{align}
where $P_{Mn}$ operates on the $\rho_{mn}^{(j)}(x,y)$ derived from $f_n(x,y,z)$, and 
\begin{align}
P_M \mathbf{f}^{(j)}
&=
\Big\{ P_{Mn} \; \rho_{mn}^{(j)}(x,y) \Big\}
\label{eq_intensity_data_proj}
\end{align}
is the result of operating on all $N$ $f_n(x,y,z)$.
Equation~(\ref{eq_intensity_data_proj}) will make the least amount of change to the set $\{ \rho_{mn}^{(j)}(x,y) \}$ in the Euclidean sense such that the views calculated from the output of $P_M$ will equal the measured views $I_n^{\;\mathrm{data}}(x,y)$ for all $n$ given any input $\rho_{mn}^{(j)}(x,y)$.  $P_M$ may be understood as a ``data satisfaction'' projection operator that generates 3D models that agree with the measured 2D views (one model for each view). This kind of problem is analogous to the situation studied by \cite{Chen_etal_2016} where an object is reconstructed from the averaged diffracted intensities from a number of object ``clusters." The ``clusters" in this case are the individual sections in each 3D volume. 
The data satisfaction projection operator $P_M$ results in a set of inconsistent models that need to be addressed by an additional projection operator, which we describe next.

The second projection operator, $P_S$, consists of the following steps: (1) Inverse propagate each $\rho_{mn}^{(j)}(x,y)$ from the set $\{ \rho_{mn}^{(j)}(x,y) \}$ with the operator $\mathcal{L}_{mn}^{-1}$. 
(2) Form $N$ 3D volumes from the set of all $MN$ inverse propagated sections. (3) Inverse rotate each volume to a common orientation by $\mathcal{R}_n^{-1}$. (4) Average the $N$ volumes together (the concur step). (5) Apply any other constraints such as support, reality or positivity that may be available, collectively denoted here by the operator $\mathcal{C}$. (6) Rotate each volume such that their orientation corresponds to their respective views by $\mathcal{R}_n$, and then finally, (7) forward propagate the result with the forward propagation operator $\mathcal{L}_{mn}$ and redistribute the propagated sections to form an updated set of $\{ \rho_{mn}^{(j)}(x,y)\}$.
The projection operator $P_S$ as described by these steps can be written concisely as
\begin{equation}
P_S \mathbf{f}^{(j)}
= \mathcal{L} \; \mathcal{C}
\; \mathcal{L}^{-1} \mathbf{f}^{(j)}
\label{eq_PS_proj}
\end{equation}

\subsection{Our algorithm}
\begin{figure}[!ht]
\centering
\includegraphics[scale=0.70]{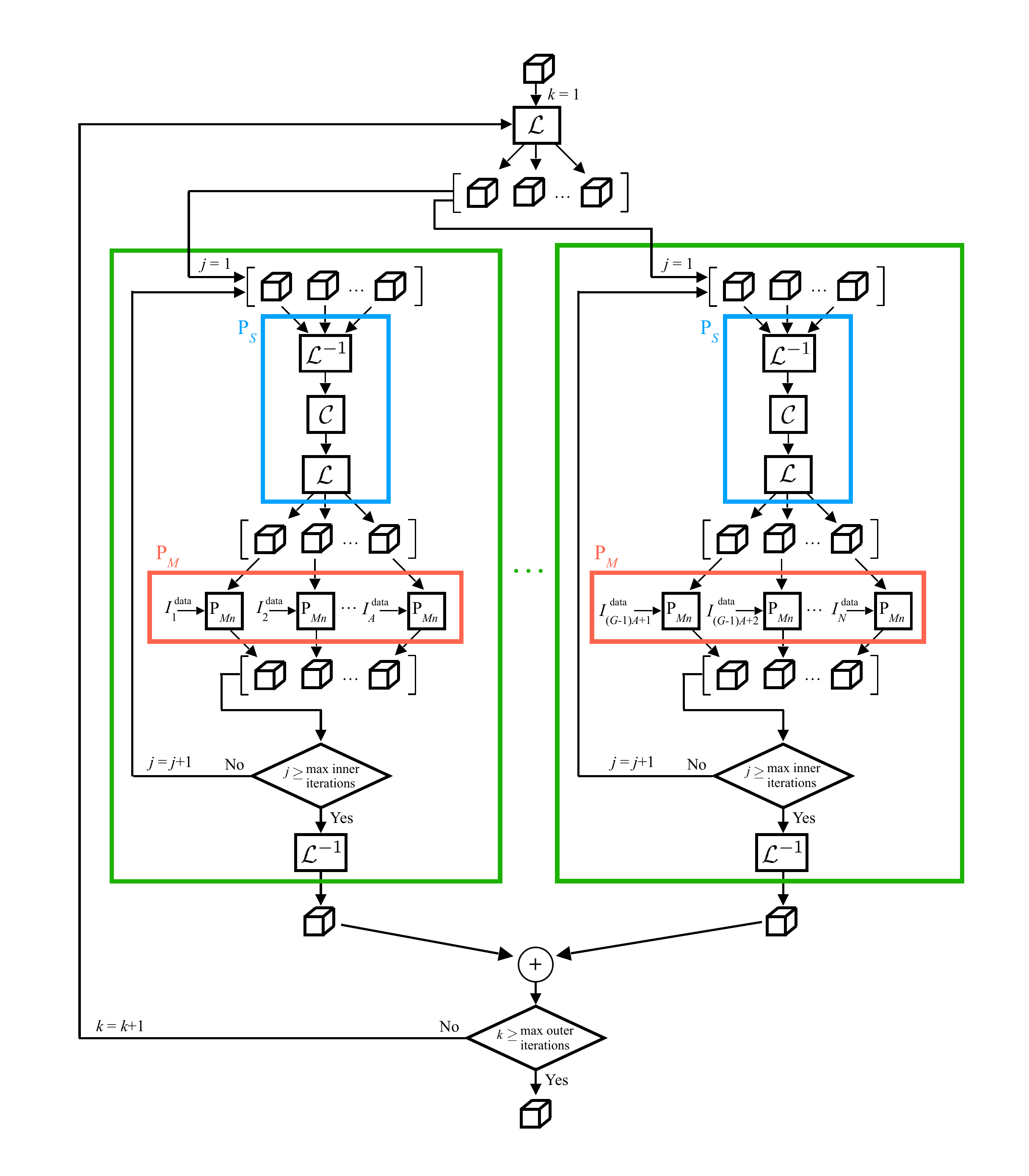}
\caption{Flow diagram of the proposed algorithm with the iterative projection algorithm chosen to be in the error reduction (ER) configuration for ease of illustration. Each green box contains a separate iterative projection algorithm, each running on a separate ``group" of data. The variable $A$ is the number of data views assigned to the first $N-1$ groups. $G$ is the total number of groups. The blue and red boxes contain the $P_S$ and $P_M$ projection operators, respectively. The iterate consists of the sets of 3D arrays shown enclosed in the square brackets.}
\label{fig_algo_flowDiagram}
\end{figure}

The number of views, $N$, is typically large in practice, thus the computer memory required to store the iterate $\mathbf{f}$ can become an issue. We have therefore structured our algorithm in a way such that the IPA is done in groups. The set of all data views $\{ I_n^{(\mathrm{data})}(x,y)\}$ is split into $G$ groups, each worked on by a separate IPA. The results from each group are averaged together after some number of IPA iterations and that averaged volume becomes the new input to each IPA in the next iteration. This procedure is repeated some number of times.
The result of this restructuring is that only approximately $N/G+1$ volumes need to be stored in memory at any given time (the +1 is for the accumulator volume). Thus the memory demand of our algorithm can be controlled by selecting the number of groups $G$.
We refer to the iterations where the output of the groups are averaged as the ``outer loop", indexed by $k$, and the iterations of the IPA itself as the ``inner loop", indexed as before by $j$. 
We denote the total number of iterations in each of these loops by $K$ and $J$, respectively.
A flow chart of our overall algorithm is shown in Figure~\ref{fig_algo_flowDiagram}.

A downside of this restructuring is that the operations no longer consist of making the least amount of change to the iterate to satisfy the set of all data views. The output of $P_M$ now satisfies the data views only in their respective groups. Only in the limit of one group ($G=1$), does the algorithm revert back to the original algorithm composed solely of proper projection operations.
However this restructuring and grouping is necessary when there could potentially be hundreds of thousands of views. Such an algorithm structure is also extremely amenable to parallelization, both for the individual groups of IPAs and also within the $P_M$ projection operator itself where each $P_{Mn}$ operation can be carried out in parallel across all $n$.

\clearpage
\section{Simulations}
\label{sec_Simulation}
We tested our algorithm on simulated 3D volumes. 
The particle that was used for the simulations was the virus capsid of a tobacco necrosis virus (protein data bank (PDB) identification: 1TNV) determined by x-ray crystallography \cite{Bando_etal_1994}. 
The virus capsid at the center of the unit cell was taken as the particle to be reconstructed with a size of approximately $L=29$nm. Since the PDB file only provided the atomic coordinates of the capsid, the interior of the virus was filled with the average electron density of the capsid for our simulations. 
The Python software package ``reborn" \cite{reborn_website} was used to convert the PDB file into a 3D array of real and positive values which we take as the scaled object potential $f(\vec{r})$.
The views were calculated via equation~(\ref{eq_I_data_L}) with the freespace propagators, (\ref{eq_propagtor_for}) and (\ref{eq_propagtor_inv}), implemented using the fast Fourier transform (FFT) algorithm.
Rotations of the 3D volume required to compute the views, i.e., equation~(\ref{eq_f2f_n_rotation}), was implemented via a series of 2D rotations of the planar slices of the 3D volume. 
Each 2D rotation was in turn performed by three shears as described by \cite{Unser_etal_1995}. 
The rotations were specified by three Euler angles, and the random rotations that were needed to generate randomly oriented views were obtained by sampling those Euler angles.

The IPA used for the reconstructions in this paper was the RAAR algorithm.
Reality and positivity constraints were applied, i.e., the operator $\mathcal{C}$, introduced in equation~(\ref{eq_PS_proj}), consists of setting the imaginary parts and the negative real parts of $f(\vec{r})$ to zero. No other additional constraints were applied, in particular, the support constraint was not used.
The starting 3D volume as input to the algorithm was filled with uniformly distributed random values between 0 and 1.
The progress of the reconstruction was monitored by calculating the errors 
\begin{equation}
e^{(j,k)} = \sqrt{ 
\frac{1}{G} \sum_{g=1}^G
 \frac{\sum_\mathbf{\vec{r}} \left( f_g^{(j,k)}(\mathbf{\vec{r}}) - f^{\mathrm{true}}(\mathbf{\vec{r}}) \right)^2}{\sum_\mathbf{\vec{r}} \left( f^{\mathrm{true}}(\mathbf{\vec{r}})\right)^2} }
\label{eq:ReconErrors_e}
\end{equation}
and
\begin{equation}
E^{(j,k)} = \sqrt{
\frac{1}{G} \sum_{g=1}^G
\frac{\sum_n \sum_{x,y} \left( I_{gn}^{(j,k)}(x,y) - I_{gn}^{\mathrm{data}}(x,y) \right)^2}{\sum_n \sum_{x,y} \left( I_{gn}^{\mathrm{data}}(x,y) \right)^2}} , 
\end{equation}
where $e^{(j,k)}$ is the root-mean-squared error between the reconstructed virus at the $j$th and $k$th iterations of the inner and outer loop, and the ground truth, $f^{\mathrm{true}}(\mathbf{\vec{r}})$, where the mean is taken over all groups $G$; similarly, $E^{(j,k)}$ is the root-mean-squared error between the set of all $N$ views generated by the iterate at the $j$th and $k$th iterations of the inner and outer loop, calculated according to equation~(\ref{eq_I_data_linop}), and the set of all data views, $\{I_n^{\mathrm{data}}(x,y)\}$, where again, the mean is taken over all groups.
The errors are appended at the beginning of the next outer loop to the errors from the previous outer loop to form the full error vectors, $e$ and $E$, at the end of each inner loop iteration, such that the full error vectors can be written as
\begin{align}
e &= [ e^{(j,1)}, \cdots, e^{(j,K)} ]
\\
E &= [ E^{(j,1)}, \cdots, E^{(j,K)} ] \;.
\end{align}

Reconstructions from two cases with data views that all have significant Ewald curvature are shown in the next two subsections. The first case is when the Ewald curvature arises due to a large incident beam wavelength, and the second case is when a high resolution is desired.
Two reconstructions are carried out separately for each case, one uses the forward and inverse propagation operators as defined in equations~(\ref{eq_op_for}) and (\ref{eq_op_inv}), while the other reconstruction has the forward and inverse propagation operators changed to
\begin{align}
\mathcal{L}_{mn} &= 
\mathcal{P}(z_n)
\\
\mathcal{L}_{mn}^{-1} &= 
\mathcal{P}^{-1}(z_n) \;,
\end{align}
i.e., with the Ewald sphere assumed flat.
The reason for doing this is to allow a comparison between an algorithm which does not take Ewald curvature into account with the proposed algorithm which does.
The quality of the two different reconstructions are gauged by the Fourier shell correlation (FSC) metric \cite{Saxton_Baumeister_1982,vanHeel_Schatz_2005}, defined as 
\begin{equation}
\mathrm{FSC}({q}) = 
\left| 
\frac{ \sum_{|\mathbf{q}|=q} F_{\mathrm{true}}(\mathbf{q})F_{\mathrm{recon}}^*(\mathbf{q}) }
{ \sqrt{\sum_{|\mathbf{q}|=q} |F_{\mathrm{true}}(\mathbf{q})|^2} 
  \sqrt{\sum_{|\mathbf{q}|=q} |F_{\mathrm{recon}}(\mathbf{q})|^2}} 
\right|,
\end{equation}
where $F_{\mathrm{true}}(\mathbf{q})$ and $F_{\mathrm{recon}}(\mathbf{q})$ are the Fourier transforms of the ground truth and the reconstructed volumes, respectively.

\subsection{Long wavelength}
\label{sec_longwavelength}
\begin{figure}[!ht]
\centering
\includegraphics[scale=0.7]{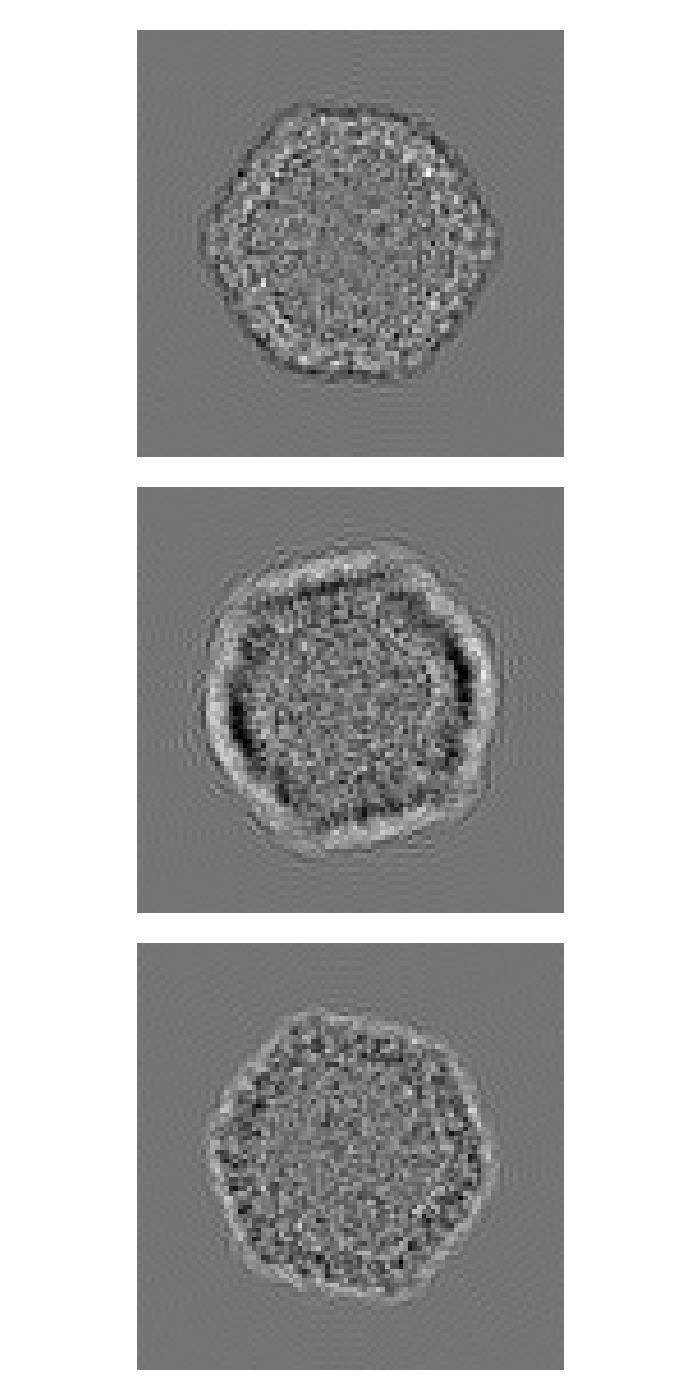}
\caption{Example data views for the case of long wavelength at three different focal distances $z_n$ and orientations $\mathcal{R}_n$ with significant Ewald curvature ($\varepsilon = 0.5$).  
The greyscales are set to the same maximum and minimum values in all subfigures.}
\label{fig_sim1_exampleViews}
\end{figure}

\begin{figure}[!ht]
\centering
\includegraphics[scale=0.75]{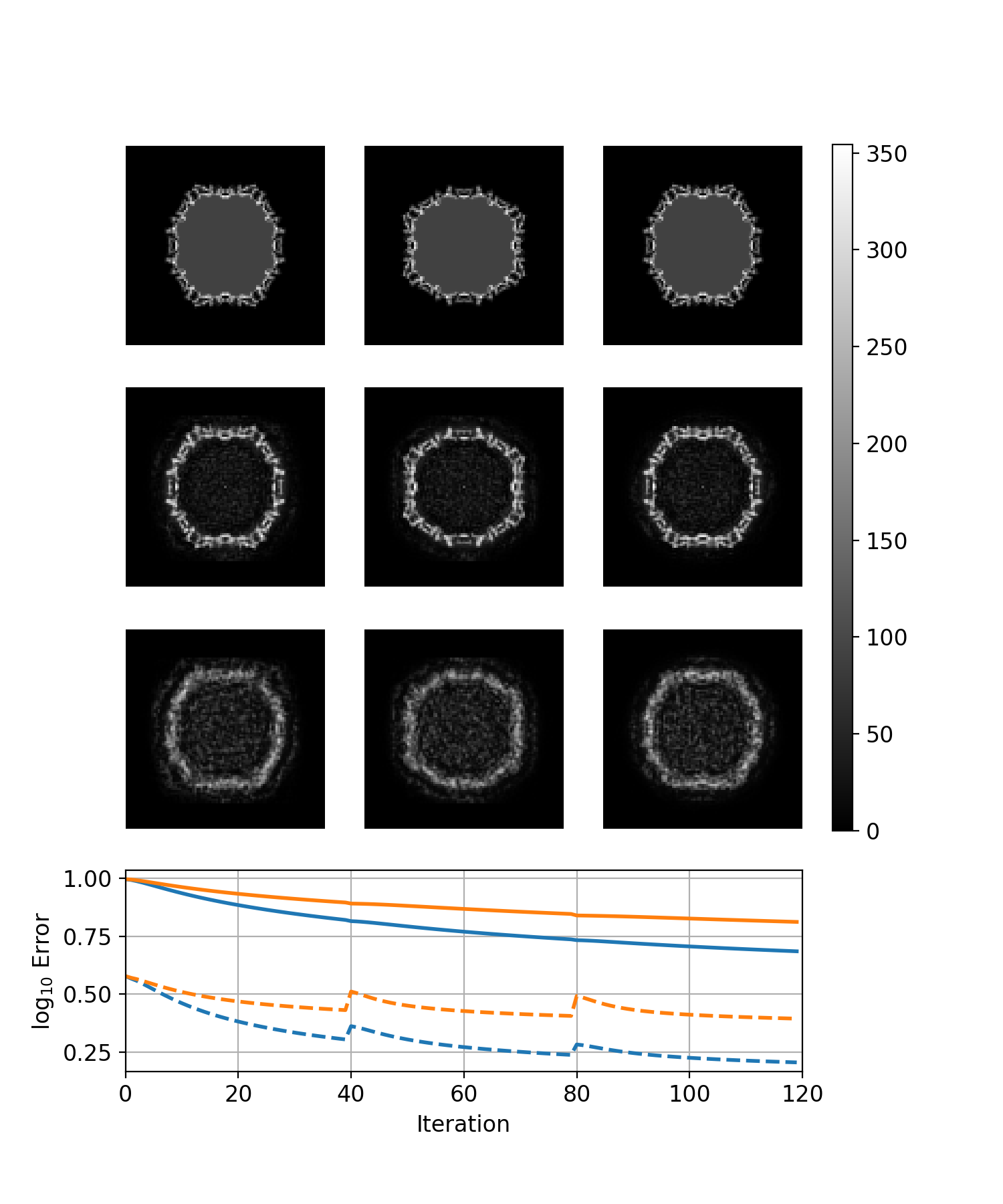}
\caption{Reconstruction results for the long wavelength case. Orthogonal central slices of the ground truth (Row 1); the reconstructed particle with the proposed algorithm (Row 2); the reconstructed particle with the proposed algorithm but ignoring Ewald curvature (Row 3). The two sets of reconstructions are scaled to have the same mean as the ground truth. The greyscale of the two reconstructions are also set to be the same as that for the ground truth. (Bottom row) Errors as the algorithm progresses on a logarithmic scale. The solid line is the object error, $e$, and the dashed line is the data error, $E$. Blue is the proposed algorithm, orange is the proposed algorithm but ignoring Ewald curvature.}
\label{fig_sim1_results}
\end{figure}

\begin{figure}[!ht]
\centering
\includegraphics[scale=0.7]{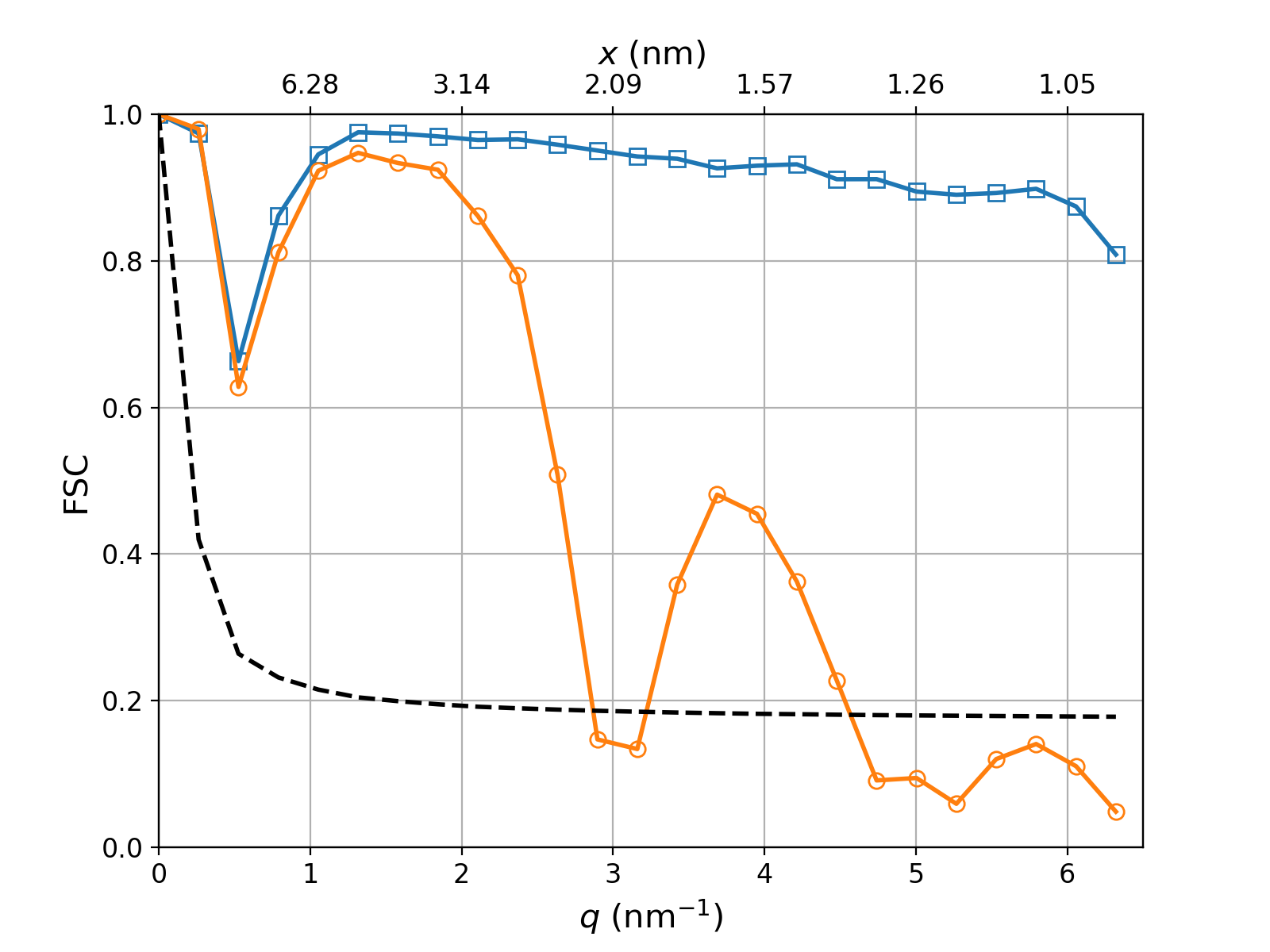}
\caption{Fourier shell correlations for the long wavelength case, comparing the ground truth virus density with the reconstruction from the proposed algorithm (blue open squares) and the reconstruction with the proposed algorithm but ignoring Ewald curvature (orange open circles). The $1/2$-bit threshold is shown as the black dashed line.}
\label{fig_sim1_results_fsc}
\end{figure}

For this first case, the virus was calculated to $d=5$~\AA{} resolution and the incident electrons have a wavelength $\lambda=0.34$~\AA{}, giving $\varepsilon = 0.5$ to one significant figure (a depth-of-field of $14.3$nm, with size of the virus $L=29$nm). 
The resultant virus is $68 \times 68 \times 68$ voxels and the computational volume is zero-padded to $93 \times 93 \times 93$ voxels.
A total of 68 views were calculated from random rotations of the virus, with randomly chosen focal distances ranging between a full length of the virus, $L$, on either side of the virus. 
Three example data views for this case of long wavelength are shown in Figure~\ref{fig_sim1_exampleViews}.
The reconstruction from this dataset is shown in Figure~\ref{fig_sim1_results} with the number of groups $G=3$, an outer loop of $K=3$ iterations, and an inner loop of $J=40$ iterations.
The RAAR algorithm parameter was set to $\beta=0.7$.
The Fourier shell correlation for the reconstructions is shown in Figure~\ref{fig_sim1_results_fsc}.

\subsection{High resolution}
\label{sec_highres}
\begin{figure}[!ht]
\centering
\includegraphics[scale=0.7]{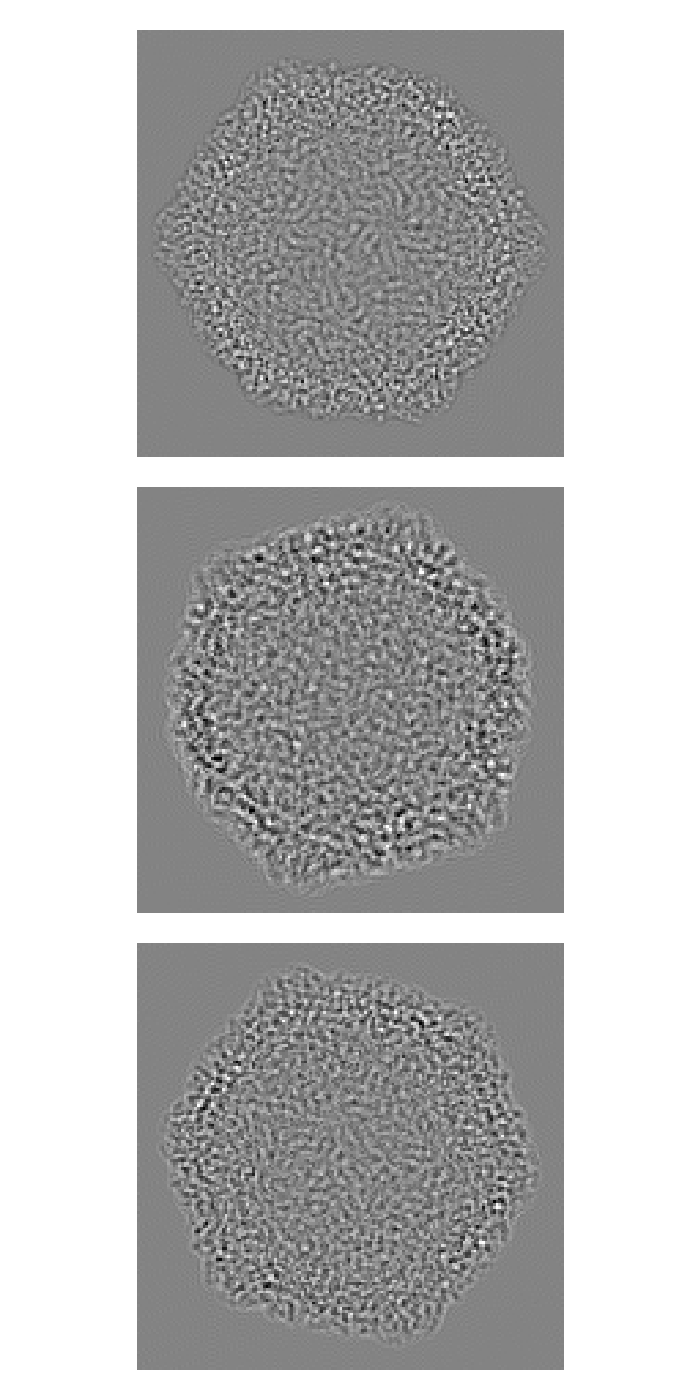}
\caption{Example data views for the case of high resolution at three different focal distances $z_n$ and orientations $\mathcal{R}_n$ with significant Ewald curvature ($\varepsilon = 0.9$).  
The greyscales are set to the same maximum and minimum values in all subfigures.}
\label{fig_sim2_exampleViews}
\end{figure}

\begin{figure}[!ht]
\centering
\includegraphics[scale=0.75]{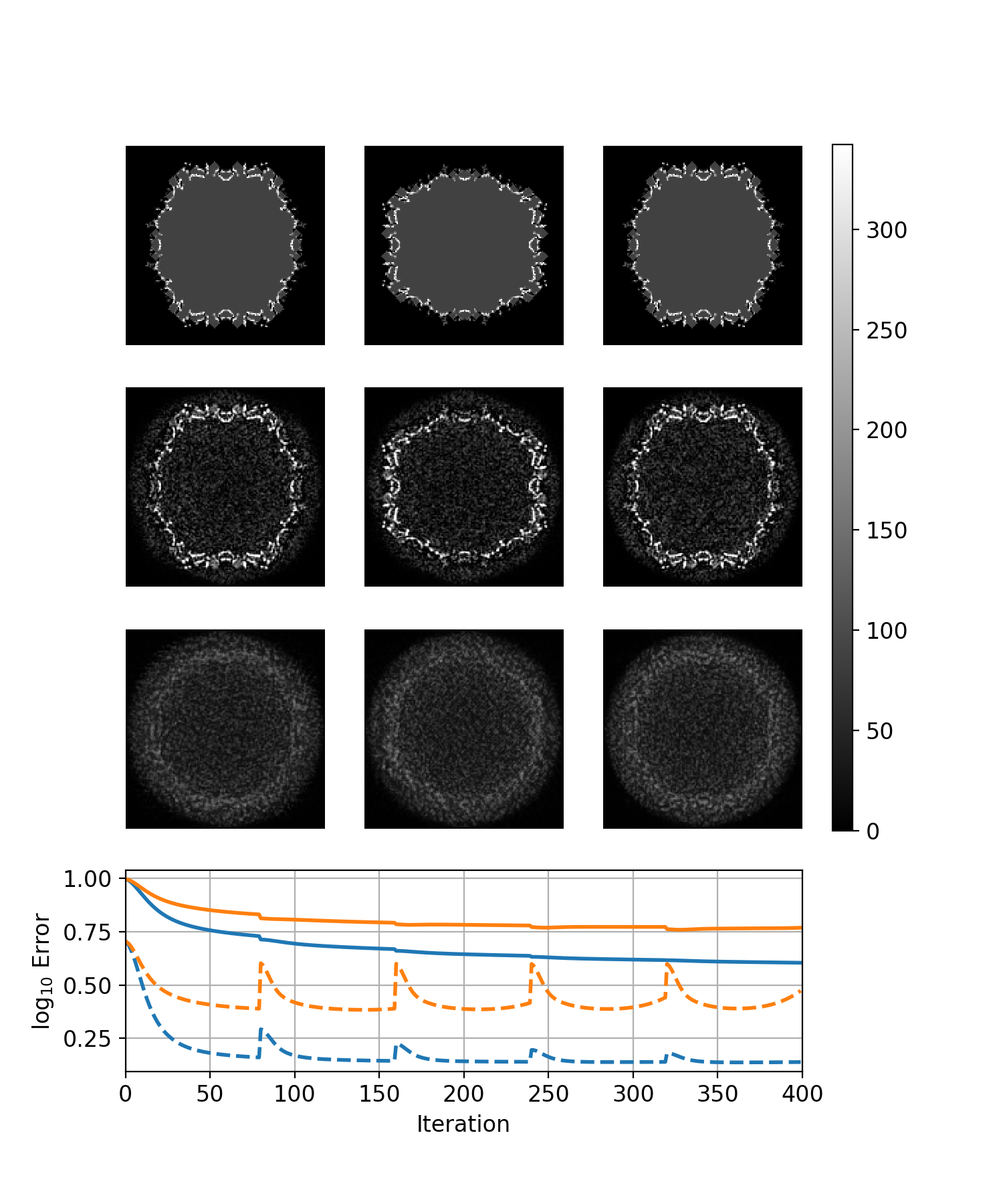}
\caption{Reconstruction results for the high resolution case. Orthogonal central slices of the ground truth (Row 1); the reconstructed particle with the proposed algorithm (Row 2); the reconstructed particle with the proposed algorithm but ignoring Ewald curvature (Row 3). The two sets of reconstructions are scaled to have the same mean as the ground truth. The greyscale of the two reconstructions are also set to be the same as that for the ground truth. (Bottom row) Errors as the algorithm progresses on a logarithmic scale. The solid line is the object error, $e$, and the dashed line is the data error, $E$. Blue is the proposed algorithm, orange is the proposed algorithm but ignoring Ewald curvature.}
\label{fig_sim2_results}
\end{figure}

\begin{figure}[!ht]
\centering
\includegraphics[scale=0.7]{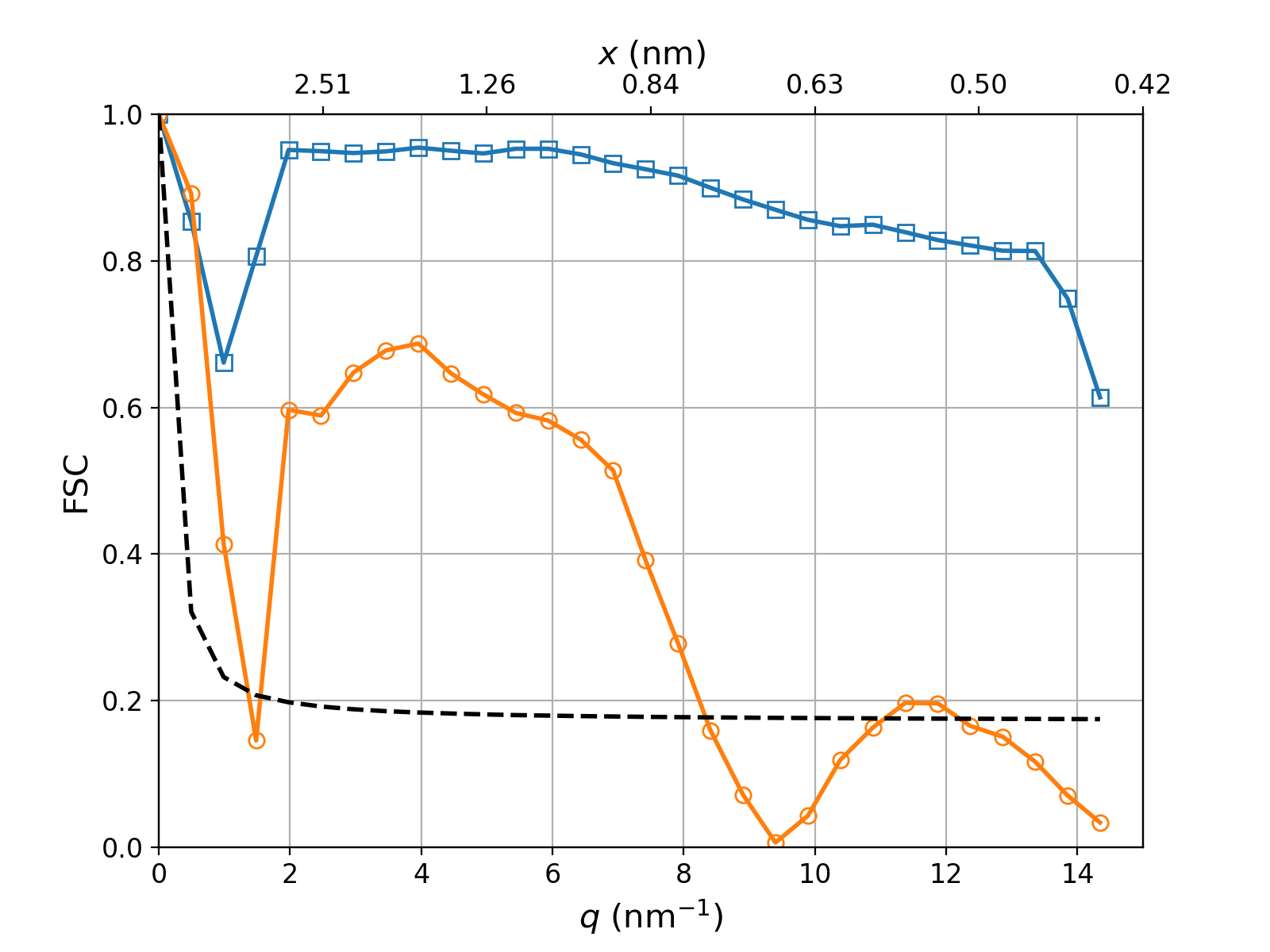}
\caption{Fourier shell correlations for the high resolution case, comparing the ground truth virus density with the reconstruction from the proposed algorithm (blue open squares) and the reconstruction with the proposed algorithm but ignoring Ewald curvature (orange open circles). The $1/2$-bit threshold is shown as the black dashed line.}
\label{fig_sim2_results_fsc}
\end{figure}

For the second case, the virus density was calculated to $d=2.2$~\AA{} resolution and the incident electrons have a wavelength $\lambda=0.0037$nm (100keV electrons), giving $\varepsilon = 0.9$ to one significant figure (a depth-of-field of 26~nm, with size of the virus $L=29$nm). 
The resultant virus is $130 \times 130 \times 130$ voxels and the computational volume is zero-padded to $156 \times 156 \times 156$ voxels.
A total of 69 views were calculated from random rotations of the virus, with randomly chosen focal distances ranging between a full length of the virus, $L$, on either side of the virus.
Three example data views for this case of high resolution are shown in Figure~\ref{fig_sim2_exampleViews}.
The reconstruction from this dataset is shown in Figure~\ref{fig_sim2_results} with the number of groups $G=2$, an outer loop of $K=5$ iterations, and an inner loop of $J=80$ iterations.
The RAAR algorithm parameter was set to $\beta=0.9$.
The Fourier shell correlation is shown in Figure~\ref{fig_sim2_results_fsc}.

\clearpage
\section{Discussion and Conclusion}
\label{sec_Conclusion}
An algorithm for reconstructing an object from projection images affected by Ewald sphere curvature in cryo-electron microscopy is proposed. 
This algorithm was shown via simulations to be able to reconstruct the correct 3D object from a set of 2D near-field intensities, which we have called ``views," that are affected by significant Ewald curvature.
A criterion for the Ewald curvature to become significant is derived, relating together the three key parameters of (1) wavelength of the incoming wave, (2) desired resolution, and (3) thickness of the sample.
Only the reality and positivity constraints are applied in our reconstructions, i.e., the object is assumed to be real and positive. A support constraint is not used.

The algorithm is based on the paradigm of iterative projection algorithms (IPAs), and a projection operation that makes the minimum change to a set of complex numbers such that the sum of their imaginary parts is equal to a desired value is derived.
A restructuring of the traditional IPA loop was proposed which alleviates the memory requirement of the algorithm when the dataset contains many views.
An implication of this restructuring is that the operations no longer consist of making the least amount of change to the iterate to satisfy the set of all data views, because the output of the projection operator $P_M$ now satisfies the data views only in their respective groups. However this is necessary for practical applications when there could potentially be hundreds of thousands of views.
Such an algorithm structure is extremely amenable to parallelization, both for the individual groups of IPAs and also for the $P_M$ projection operation.

For this proof-of-concept work, we have not parallelized the algorithm. The computational complexity of our algorithm in serial implementation is $O(N^{4} \log N)$ where $N$ is the number of voxels in each dimension of the 3D array. A factor $N^3 \log(N)$ is due to the 3D FFT, and the remaining factor of $N$ comes from the fact that the number of 2D views needed for a unique reconstruction is proportional to $N$.
For unoptimized Python code and working solely with double-precision floating-point numbers, the reconstruction detailed in Section~\ref{sec_longwavelength}, the long wavelength case, took around 19 hours. The reconstruction detailed in Section~\ref{sec_highres}, the high resolution case, took around 8.5 days. 
Both reconstructions were carried out with a single core on ASU's Agave research computing system utilizing Intel Broadwell CPUs.
Larger objects may require more iterations of the algorithm.
The compute time can likely be reduced if the initial starting iterate of the algorithm was a preliminary reconstruction obtained by treating the views as projections, i.e., ignoring Ewald curvature. 
In terms of storage requirement, our code stores the set of all $\rho_{mn}(x,y)$ as the iterate $\mathbf{f}$. If we switch to storing the set of all $f_n(x,y,z)$ then that would further reduce the amount of storage needed. The tradeoff is time, because $\rho_{mn}(x,y)$ will have to be generated from $f_n(x,y,z)$ by zero-padding and Fourier transforming at every iteration. 
As mentioned before, the greatest speed-up would come from parallelizing the algorithm, which we are working towards.

In the limit where the Ewald sphere is flat, our algorithm becomes a tomographic reconstruction method, capable of recovering a 3D object from conventional tomographic data where each view is just a simple projection (sum of densities) through the object.

Our method assumes the use of a conventional through-focus series for image reconstruction. In comparison with an alternative method based on the far-out-of-focus spatial separation of spatial frequencies \cite{Russo_2018}, while our approach lacks the benefits of masking in that method, it may have particular advantages when used with a Zernike phase plate \cite{Danev_Baumeister_2017}. Using in-focus images, this preserves low spatial frequencies in image formation, otherwise lost in bright-field out-of-focus images of a weak phase object. Phase plates thus avoid the need for the very large defocus (with loss of high-resolution detail) required to obtain visibility from the smallest particles, so that our method could then give access to these smaller biomolecules in this way. It may also reduce damage by reducing the number of image recordings needed. Our approach is more computationally intensive, but should be simpler to implement experimentally.

Our algorithm assumes we know the orientation and defocus of the object in each view, both of which should be obtainable from existing cryo-EM software. For cryo-EM tomography, the resolution 
limit imposed by radiation damage means that Ewald sphere curvature is 
unlikely to be important, so that we have assumed that this method will be 
applied to single-particle data. Additional analysis to allow for 
conformation variation may be considered in future developments of this method.
The contrast transfer function (CTF) is ignored here in this work. In the event that the CTF is known, and is assumed to be the same for all views, then it can be incorporated into the current algorithm and does not require change to the overall structure of the method.
If the CTF is unknown, then the lens aberration parameters may be recoverable in addition to the object potential, using a modified version of the proposed algorithm, a line of inquiry we are currently investigating.
Further future work includes exploring the effect of varying some of the parameters of the algorithm, such as the number of groups, $G$, the max inner iteration, $J$, and the max outer iteration, $K$, and extending the algorithm to deal with multiple scattering.

The code for the simulations carried out in this paper can be found at: 
https://gitlab.com/jpchen1/em-reconstruction-with-ewald
\\

\textbf{Acknowledgements}: We are grateful to Prof R.~M.~Glaeser for useful conversations in connection with this work during his visit to the new Cryo-EM center at ASU.
We also thank Cornelius Gati for many helpful discussions and the two anonymous referees for their encouragement and many constructive comments.
We acknowledge the ASU Agave computing cluster and NVIDIA Corporation for their Titan V GPU.
JPJC and RAK acknowledge support from NSF STC Award DBI-1231306. JPJC, KES and RAK acknowledge support from NSF Award DBI-1565180. JCHS acknowledge support from ARO award  AWD00035320.

\appendix

\section{Potential scattering}
\label{appendix_LS}
Assuming non-relativistic elastic scattering, we start with the time-independent Schr\"odinger equation
\begin{align}
 \frac{-\hbar^2 }{2m}\nabla^2 \psi(\vec{r}) - eU(\vec{r}) \psi(\vec{r}) = E \psi(\vec{r}) \;,
\end{align}
where $U(\vec{r})$ is the electric potential, $e$, $m$ and $E$ are the charge, mass and kinetic energy of the electron, respectively. 
Define $k_0 = \sqrt{2mE}/\hbar$ and $V(\vec{r}) = -e U(\vec{r})$ we can write down the inhomogeneous Helmholtz's equation
\begin{align}
 (\nabla^2 + k_0^2) \psi(\vec{r}) =  \frac{2m}{\hbar^2} V(\vec{r}) \psi(\vec{r}) \;.
\label{eq_A0}
\end{align}
The integral form of equation~(\ref{eq_A0}) is the Lippmann-Schwinger equation, which, assuming an incoming plane wave $e^{ik_0 z}$, can be written as 
\begin{equation}
\psi(\vec r) = e^{ik_0 z} 
+ \frac{2m}{\hbar^2} \int d^3r'
G(\vec{r}, \vec{r}') V(\vec r') \psi(\vec r')
\,,
\label{eq_A0_2}
\end{equation}
where $G(\vec{r}, \vec{r}')$ is the Green's function of the homogenous Helmholtz's equation,
\begin{equation}
G(\vec{r}, \vec{r}')
=
- \frac{e^{ik_0|\vec r- \vec r'|}}{4\pi |\vec r -\vec r'|} \;.
\end{equation}
Making the first Born approximation and factoring out the $e^{ik_0z}$ term in equation~(\ref{eq_A0_2}), gives the scattered wave
\begin{equation}
\psi(\vec r) = e^{ik_0 z}\left (
1 
+ \frac{2m}{\hbar^2} \int d^3r'
G(\vec{r}, \vec{r}') V(\vec r') e^{-ik_0 (z- z')}
\right )
\label{eq_A1}
\end{equation}
as the solution to our scattering problem.
To cast this solution into the form of freespace propagators, first write the Helmholtz Green's function as
\begin{equation}
G(\vec{r}, \vec{r}')
=
\int \frac{d^3k}{(2\pi)^3}
\frac{ e^{i\vec k \cdot (\vec r-\vec r')}}
{k_0^2-k^2 + i \eta} \,,
\label{eq_A3_HelmhotzGreenFuncInt}
\end{equation}
where $\eta$ is a positive infinitesimal.
Evaluating the $k_z$ integral in equation~(\ref{eq_A3_HelmhotzGreenFuncInt}) gives,
\begin{equation}
G(\vec{r}, \vec{r}')
 = 
-i\int \frac{dk_x}{2\pi} \int \frac{dk_y}{2\pi}
e^{ik_x (x-x')}
e^{ik_y (y-y')}
\frac{e^{i\sqrt{k_0^2-k_x^2-k_y^2}|z-z'|}}
{2\sqrt{k_0^2-k_x^2-k_y^2}}
\,,
\label{eq_A4}
\end{equation}
where the square root is defined to give decaying exponentials if $k_x^2+k_y^2 > k_0^2$.
Substituting equation~(\ref{eq_A4}) into equation~(\ref{eq_A1}),
taking $z > z'$,
and replacing the square root in the denominator by $k_0$ due to the small angle scattering geometry typical in electron microscopy experiments gives our final result
\begin{equation}
\psi(\vec r) = e^{ik_0 z} \left [ 1
-\frac{mi}{\hbar^2 k_0} \int d^3r'
\int \frac{dk_x}{2\pi} \int \frac{dk_y}{2\pi}
e^{ik_x (x-x')}
e^{ik_y (y-y')}
e^{i\left (\sqrt{k_0^2-k_x^2-k_y^2}-k_0\right)(z-z')}
V(x',y',z') \right ] \,.
\end{equation}
Writing $E =\frac{\hbar^2 k_0^2}{2m}$ and $\lambda = \frac{2\pi}{k_0}$,
the prefactor of the integral can be written as
$\frac{mi}{\hbar^2 k_0} = i\frac{\pi}{\lambda E}$.

\section{Projection operator for constraining the sum of imaginary numbers}
\label{appendix_Im_proj}
In this appendix we derive the operation that makes the minimum change to a set of complex numbers such that the sum of their imaginary parts becomes equal to a desired value. 
Denote the desired value by $I^{\mathrm{data}}$, and the original set of $N$ complex numbers by $z_n^o = x_n^o + i y_n^o$. The sum of all imaginary parts from the $N$ complex numbers is then
\begin{align}
I^o = \sum_{n=1}^{N} y_n^o \;.
\end{align}
We would like the set of complex numbers after the projection operation, denoted by $z_n = x_n + i y_n$, to have $\sum_{n=1}^{N} y_n = I^{\mathrm{data}}$.

The function we wish to minimize is the Euclidean distance
\begin{align}
f = \sum_{n=1}^{N} \big[ (x_n - x_n^o)^2 + (y_n - y_n^o)^2 \big] \;.
\end{align}
The constraint equation is
\begin{align}
g = \sum_{n=1}^{N} y_n - I^{\mathrm{data}} \;.
\end{align}
Applying the method of Lagrange multipliers leads us to write
\begin{align}
L = f + \lambda g \;,
\end{align}
where $\lambda$ is the Lagrange multiplier.
Taking the partial derivatives of $L$ with respect to $x_n$ and $y_n$ and setting them to zero gives
\begin{align}
0 &= \frac{\partial L}{\partial x_n} = 2(x_n - x_n^o)
\\
0 &= \frac{\partial L}{\partial y_n} = 2(y_n - y_n^o) + \lambda \;,
\end{align}
yielding
\begin{align}
x_n &= x_n^o
\label{eq_Append2_xn}
\\
y_n &= y_n^o - \frac{\lambda}{2} \;.
\label{eq_Append2_yn}
\end{align}
Summing over all $n$ in equation~(\ref{eq_Append2_yn}) and rearranging for $\lambda$ gives
\begin{align}
\lambda = \frac{2}{N} \sum_{n=1}^{N} \big( y_n^o - y_n \big) \;.
\label{eq_Append2_lambda}
\end{align}
Substituting equation~(\ref{eq_Append2_lambda}) back into (\ref{eq_Append2_yn}) and combining the real and imaginary parts using (\ref{eq_Append2_xn}) finally yields
\begin{align}
z_n = z_n^o + i\frac{1}{N} \left( I^{\mathrm{data}} - I^o \right) \;.
\label{eq_Append2_zn_final}
\end{align}
Equation~(\ref{eq_Append2_zn_final}) is the projection operation that makes the minimum change to a set of complex numbers, measured by the Euclidean distance, such that the sum of their imaginary parts is equal to $I^{\mathrm{data}}$.

\bibliography{paper1}

\end{document}